\documentstyle[12pt]{article}
\newcommand{\im}{\mathop{\Im m}\nolimits}
\renewcommand{\O}{{\cal O}}

\catcode`@=11
\newcount\@tempcntc
\def\@citex[#1]#2{\if@filesw\immediate\write\@auxout{\string\citation{#2}}\fi
  \@tempcnta\z@\@tempcntb\m@ne\def\@citea{}\@cite{\@for\@citeb:=#2\do
    {\@ifundefined
       {b@\@citeb}{\@citeo\@tempcntb\m@ne\@citea\def\@citea{,}{\bf ?}\@warning
       {Citation `\@citeb' on page \thepage \space undefined}}%
    {\setbox\z@\hbox{\global\@tempcntc0\csname b@\@citeb\endcsname\relax}%
     \ifnum\@tempcntc=\z@ \@citeo\@tempcntb\m@ne
       \@citea\def\@citea{,}\hbox{\csname b@\@citeb\endcsname}%
     \else
      \advance\@tempcntb\@ne
      \ifnum\@tempcntb=\@tempcntc
      \else\advance\@tempcntb\m@ne\@citeo
      \@tempcnta\@tempcntc\@tempcntb\@tempcntc\fi\fi}}\@citeo}{#1}}
\def\@citeo{\ifnum\@tempcnta>\@tempcntb\else\@citea\def\@citea{,}%
  \ifnum\@tempcnta=\@tempcntb\the\@tempcnta\else
   {\advance\@tempcnta\@ne\ifnum\@tempcnta=\@tempcntb \else \def\@citea{--}\fi
    \advance\@tempcnta\m@ne\the\@tempcnta\@citea\the\@tempcntb}\fi\fi}
\catcode`@=12
\begin{document}
\title{\vskip-3cm{\baselineskip16pt
\centerline{\normalsize DESY 93-194\hfill ISSN 0418-9833}
\centerline{\normalsize NYU-Th-93/12/01\hfill}
\centerline{\normalsize hep-ph/9401243\hfill}
\centerline{\normalsize December 1993\hfill}}
\vskip1.5cm
Observations Concerning the Magnitude of $t \bar{t}$
Threshold Effects on Electroweak Parameters}
\author{Bernd A. Kniehl\\
II. Institut f\"ur Theoretische Physik, Universit\"at Hamburg\\
Luruper Chaussee 149, 22761 Hamburg, Germany\\ \\
Alberto Sirlin\\
Department of Physics, New York University\\
4 Washington Place, New York, NY 10003, USA}
\date{}
\maketitle
\begin{abstract}
We discuss a recent analysis of $t\bar{t}$ threshold effects and
its implications for the determination of electroweak parameters.
We show that the new formulation, when applied to the $\rho$
parameter for $m_t\approx158$~GeV, gives a result of similar magnitude to those
previously obtained.  In fact, it is quite close to our ``resonance''
calculation.  We also present a simple estimate of the size of the
threshold effects based on an elementary Bohr-atom model of
$t\bar{t}$ resonances.
\end{abstract}

\section{Introduction}

The analysis of threshold effects involving heavy quarks and their
contribution to the determination of electroweak parameters has
been the subject of a number of studies in the past
\cite{fad,kwo,str,kue,kni}.
Very recently, F.J. Yndur\'ain has discussed, in the framework of the
Coulombic approximation, $t\bar{t}$ threshold effects on the
renormalized vacuum-polarization function $\Pi(s)-\Pi(0)$ associated
with conserved vector currents \cite{ynd}.  Here $\Pi(s)$ is the
unrenormalized function defined according to
$\Pi_{\mu\nu}^V(q)=(q^2g_{\mu\nu}-q_\mu q_\nu)\Pi(q^2)$, with
$s=q^2$.  For $m_t=\sqrt3m_Z\approx158$~GeV and $s=m_Z^2$, the author
of Ref.~\cite{ynd} finds that the threshold effects are significantly
smaller than the perturbative $\O(\alpha_s)$ calculation. From this
observation he concludes that threshold effects are generally small
for $s\ll4m_t^2$ and that the ``large'' results reported in Ref.~\cite{kni}
concerning their contribution to electroweak parameters are not
supported by his ``detailed, rigorous calculation.''
However, $t\bar t$ threshold effects influence electroweak parameters chiefly
through the $\rho$ parameter.
The function $\Pi(s)-\Pi(0)$, although very important in its own right, has
very little to do with the effects discussed in Ref.~\cite{kni}.  In fact,
it is well known that heavy particles of mass $m^2\gg s$ decouple in this
amplitude.  For this reason, it should be obvious that the leading
effects discussed in Ref.~\cite{kni} do not arise from this amplitude.
Thus, conclusions drawn on the work of Ref.~\cite{kni} from the discussion
of $\Pi(s)-\Pi(0)$ are without foundation.  In order to show in the simplest
possible way what the correct conclusions ought to be, in Section~2
we apply the formulation of Ref.~\cite{ynd} to the study of leading
threshold contributions to the $\rho$ parameter for $m_t \approx 158$~GeV
and other values of $m_t$.
In principle, this requires only minor modifications of the relevant formulae
of Ref.~\cite{ynd}, mainly in the prefactors.
However, for reasons explained in that Section, we find it necessary to
re-evaluate the threshold corrections of that paper.  We then find
that, contrary to the conclusions of Ref.~\cite{ynd}, this leads to
results similar in magnitude to those reported in Ref.~\cite{kni}.  In
fact, the approach of Ref.~\cite{ynd} gives threshold corrections to the
$\rho$ parameter quite close to our own ``resonance'' calculation.
In order to make more transparent the size of these leading
threshold corrections relative to the perturbative $\O(\alpha_s)$
calculations, in Section 3 we present a simple estimate based on
an elementary Bohr-atom model of toponia.  We also briefly comment
on the shifts induced in electroweak parameters and the magnitude
of $\O(\alpha_s^2)$ corrections.

\section{Threshold corrections to the $\rho$ parameter}

For large $m_t$, the dominant threshold effects discussed in Ref.~\cite{kni}
can be related to corrections to the $\rho$ parameter.  As
the current $\bar\psi_t\gamma^\mu\psi_t$ is conserved and the $\rho$ parameter
is defined at $s=0$, it is clear that $t\bar{t}$ threshold effects arise in
this case from the contributions of the axial-vector current.  As
explained in Ref.~\cite{kni}, on account of the Ward identities such
contributions involve $\im\lambda^A(s,m_t,m_t)$,
where $\lambda^A$ is a longitudinal part of
the axial-vector polarization tensor $\Pi_{\mu\nu}^A$.  Furthermore, for
non-relativistic, spin-independent QCD potentials, $-\im\lambda^A$
can be identified to good approximation with $\im\Pi$ (in Ref.~\cite{kni},
$\Pi$ is called $\Pi^V(s)/s$).
In particular, we note that both amplitudes receive
contributions from $nS$ states.  One then finds that, in the
formulation Ref.~\cite{kni}, the leading threshold correction to the
$\rho$ parameter is given by (cf.\ Eqs.~(5.1, 5.2a) of Ref.~\cite{kni})
\begin{equation}
\label{one}
\delta(\Delta\rho)_{thr}=-{G_F\over2\pi\sqrt2}\int ds^\prime\,
\im\Pi_{thr}(s^\prime).
\end{equation}
Here $\im \Pi_{thr}(s')$ denotes contributions from the threshold region not
taken into account in the usual perturbative $\O(\alpha_s)$ calculation.

Our aim is to employ the analysis of Ref.~\cite{ynd} to calculate
Eq.~(\ref{one}) and then to compare the answer with our results.
In the formulation of that paper, based on the Coulombic approximation,
there are two contributions to $\im\Pi_{thr}(s^\prime)$.
One of them arises from the toponium resonances and is called
$\im\Pi_{pole}(s^\prime)$.  The other represents a summation of
$(\alpha_s/v)^n\ (n=1,2,\ldots)$ terms,
integrated over a small range above threshold, after
subtracting corresponding $\O(\alpha_s)$ contributions.
The first one can be obtained from the expression \cite{ynd}
\begin{equation}
\label{two}
\im\Pi_{pole}(s)=N_c\sum_n\delta(s-M_n^2)
{\left|\tilde R_{n0}^{(0)}(0)\right|^2\over M_n}
\left[1+{3\beta_0\alpha_s\over2\pi}\left(\ln{n\mu\over C_F\alpha_s m_t}
+\psi(n+1)-1\right)\right],
\end{equation}
where $N_c=3$, $C_F=(N_c^2-1)/(2N_c)=4/3$, $n_f=5$, $\beta_0=11-2n_f/3=23/3$,
$M_n$ is the mass of the $nS$ toponium resonance in a Coulombic potential,
$\left|\tilde R_{n0}^{(0)}(0)\right|^2=C_F^3\tilde\alpha_s^3(\mu)m_t^3/(2n^3)$
is the square of its radial wave function at the origin, and
$\tilde\alpha_s(\mu)=\alpha_s(\mu)(1+b\alpha_s(\mu)/\pi)$, with
$b=\gamma_E(11N_c-2n_f)/6+(31N_c-10n_f)/36\approx3.407$ for toponia
\cite{ren}.
Inserting Eq.~(\ref{two}) into Eq.~(\ref{one}), we obtain
\begin{eqnarray}
\label{three}
\delta(\Delta\rho)_{pole}&=&-x_t\pi\zeta(3)C_F^3\tilde\alpha_s^3(\mu)
\left[1+{3\beta_0\alpha_s\over2\pi}\left(\ln{\mu\over C_F\alpha_sm_t}
-\gamma_E\right)\right. \nonumber\\
& &+\left.{3\beta_0\alpha_s\over2\pi\zeta(3)}\sum_{n=2}^\infty{1\over n^3}
\left(\ln n+\sum_{k=2}^\infty{1\over k}\right)\right],
\end{eqnarray}
where $x_t=(N_cG_Fm_t^2/8\pi^2\sqrt2\,)$ is the Veltman correction to the
$\rho$ parameter \cite{vel}.
In the numerical evaluation of the threshold corrections to
$\Pi(s)-\Pi(0)$ reported in Ref.~\cite{ynd}, $\mu$ is chosen to be $m_Z$,
$m_t=\sqrt3m_Z$ is assumed, and $\alpha_s(m_Z)=0.115\pm0.01$ is taken.
Unless stated otherwise, we shall adopt these values in the following.
Furthermore,
$(15/16)\pi\zeta(3)C_F^3\tilde\alpha_s^3(\mu)$ is found to equal
$1.81\times10^{-2}$, the summation $\sum_{n=2}^\infty$ is neglected,
and $\{1+(3\beta_0\alpha_s/2\pi)[\ln(\mu/C_F\alpha_sm_t)-\gamma_E]\}$
is given as 1.08.
It is apparent that the numerical value given in
Ref.~\cite{ynd} for the last factor is too low.  The correct value is
1.315.  Inclusion of the neglected sum raises the expression
between square brackets in Eq.~(\ref{three}) to 1.437, instead of 1.08
\cite{pol}. This leads to
\begin{eqnarray}
\delta(\Delta\rho)_{pole}&=&-{16\over15}\,1.816\times10^{-2}\cdot1.437\,x_t
\nonumber\\
\label{four}
&=&-0.0278\,x_t.
\end{eqnarray}

The contribution from the small range above threshold can be
gleaned from Ref.~\cite{ynd}.  Using $v=(1-4m_t^2/s)^{1/2}$ as integration
variable ($v$ is the top-quark velocity in the center-of-mass frame)
and approximating $(1-v^2)^{-2}\approx1$ in the
integrand, the contribution to $\Delta\rho$ equals
$-(16/15)\delta_{thr}x_t$, where $\delta_{thr}$ is a quantity studied in
Ref.~\cite{ynd}.
For $s\ll4m_t^2$, $\delta_{thr}$ can be written as
\begin{eqnarray}
\label{fivea}
\delta_{thr}&=&{15\over4}\int_0^{v_0}dv\,v
\left[{B(v)\over1-e^{-B(v)/v}}-v-{\pi\over2}C_F\alpha_s(m_t)\right],\\
\label{fiveb}
B(v)&=&\pi C_F\alpha_s(\mu)\left\{1+{\alpha_s\over\pi}\left[b
+{\beta_0\over2}\left(\ln{C_F\alpha_s\mu\over4m_tv^2}-1\right)\right]\right\},
\end{eqnarray}
where an overall factor $(1-v^2/3)$ has been omitted under the integral.
In Eq.~(\ref{fivea}), the term involving $B(v)$ represents a summation of
$(\pi C_F\alpha_s/v)^n$
contributions valid for large values of this parameter, while the
last two correspond to the subtraction of the perturbative calculation
up to $\O(\alpha_s)$ in the small-$v$ limit.
We have evaluated the latter at $m_t$, as this is demonstrably the proper scale
to be employed in the perturbative calculation \cite{kni}.
We note that the integrand of Eq.~(\ref{fivea}) is renormalization-group
invariant through $\O(\alpha_s^2)$ \cite{sca}.
In Ref.~\cite{ynd}, the value $v_0=\pi C_F\alpha_s(m_t)/\sqrt2\approx0.314$
is chosen, the exponential and terms of higher order in $\alpha_s$ are
neglected, and the answer given as
\begin{equation}
\label{six}
\delta_{thr}={15\over16}\pi^3C_F^3\alpha_s^2(m_t)
\left({1\over2}\alpha_s(\mu)-{\sqrt2\over3}\alpha_s(m_t)
+{b\over\pi}\alpha_s^2(\mu)
+{\beta_0\over2\pi}\alpha_s^2(\mu)\ln{\mu\over2\pi^2C_F\alpha_sm_t}\right).
\end{equation}
Although the analytic summation of $(\pi C_F\alpha_s/v)^n$ terms is
theoretically interesting, there are unfortunately a number of problems in
the evaluation of $\delta_{thr}$ carried out in Ref.~\cite{ynd}:
\begin{enumerate}
\item Evaluation of Eq.~(\ref{six}) as it stands gives a negative result,
$\delta_{thr}=-3.82\times10^{-3}$ \cite{wro}, which obviously contradicts
the well-known fact that multi-gluon exchanges lead to an enhancement of the
$t\bar t$ excitation curve.
\item Equation~(\ref{six}) is not renormalization-group invariant,
as the coefficient of $\alpha_s(\mu)$ does not match correctly that
of $\ln\mu$. Subject to the approximations explained before, the correct,
renormalization-group invariant expression is obtained
by including an additional term $[\alpha_s(\mu)-\alpha_s(m_t)]/2$
within the parentheses of Eq.~(\ref{six}). We note in passing that
this additional term may be traced to a change of scale in the last
term of Eq.~(\ref{fivea}), from the arbitrary value $\mu$ to the
proper physical choice $m_t$.
For $\mu=m_Z$, the value of the corrected expression is
$\delta_{thr}=-3.83\times10^{-4}$, i.e., essentially zero, and very
different from the value reported in Ref.~\cite{ynd}.
\item As we have a
near cancellation of relatively large contributions and, moreover,
the result should be positive, it is clear that, for the chosen
value of $v_0$, the neglect of the exponential in Eq.~(\ref{fivea}) is not
justified.  In fact, evaluating numerically Eq.~(\ref{fivea}) with the
exponential included, we find $\delta_{thr}=1.55\times10^{-2}$.
This value is positive, as it should be, and much larger than
the answer obtained without the exponential. Actually,
it is 2.9 times larger than the result reported in Ref.~\cite{ynd}.
\end{enumerate}
\noindent
We also note that the integrand of Eq.~(\ref{fivea}) vanishes at
$1.002\,v_0$, where $v_0$ is defined above Eq.~(\ref{six}).
Thus, $v_0$ is a reasonable value to use as the upper limit of integration
because at that point the resummed series and the perturbative $\O(\alpha_s)$
contributions nearly coincide; evaluation of the integrals up to
the point where the integrand actually vanishes leads to negligible
changes.  On the other hand, a possible weakness of the method is
that $v_0$ is rather large and for such values of $v$ it is not
clear that the resummed expression is valid.  Nevertheless, for our
present purpose, which is the evaluation of the contribution to
$\Delta\rho$ according to the prescriptions of Ref.~\cite{ynd},
we use the above value of $v_0$.
For the same reason, we use $\mu=m_Z$, although this is not a
characteristic scale in connection with the $\rho$ parameter;
using the appropriate value, $\mu=m_t$, would make the radiative correction
and the overall result slightly smaller.
However, we include the effect of an additional overall factor
$(1-v^2/3)/(1-v^2)^2$,
which should be appended to the integrand of Eq.~(\ref{fivea}) in the
case of $\Delta\rho$; it
increases the result by about 4.3\%.  We then find that the contribution
to $\Delta\rho$ from the range $0\le v\le v_0$ above threshold is
$-(16/15)1.55\times10^{-2}\cdot1.043\,x_t=-0.0172\,x_t$.
Combining this result with Eq.~(\ref{four}), we find that,
after correcting the errors discussed above, the formulation of
Ref.~\cite{ynd} leads to $\delta(\Delta\rho)=-0.0450\,x_t$.
For comparative purposes, we rescale this result to the case
$\alpha_s(m_Z)=0.118$, the
value used in our calculations, and obtain, for $m_t=\sqrt3m_Z$,
\begin{equation}
\label{seven}
\delta(\Delta\rho)_{thr}= -0.0486\,x_t
\qquad\mbox{(Ref.~\cite{ynd})}.
\end{equation}

In our own work we have applied two different methods to evaluate
the imaginary parts near threshold.
The first one is the resonance approach of Ref.~\cite{kue}, which
assumes the existence of narrow, discrete $t\bar t$ bound states
characterized by $R_n(0)$ and $M_n$.
Moreover, in Ref.~\cite{kue} a specific interpolation procedure is
developed to implement the matching of the higher resonances and the
continuum evaluated perturbatively to $\O(\alpha_s)$.
The second one is the Green-function (G.F.) approach of Ref.~\cite{str},
which takes into account the smearing of the resonances by the weak decay of
its constituents and leads to a continuous excitation curve.
Both approaches make use of realistic QCD potentials, the Richardson
and the Igi-Ono potentials, which reproduce accurately charmonium and
bottonium spectroscopy and are expected to describe well toponia, too.
These QCD potentials contain a term linear in the inter-quark distance,
$r$, to account for the confinement of color.
Detailed studies reveal that the shape of the Green function is not very
sensitive to the long-distance behaviour of the potential \cite{exc}.
This may be understood by observing that the top quarks decay before they
are able to reach large distances.
The rapid weak decay of the top quarks,
which causes the screening of the long-distance effects,
is properly taken into account in the Green-function approach of
Ref.~\cite{kni}, while it is not implemented in the resonance approaches of
Refs.~\cite{kni} and \cite{ynd}.
In the latter case, it is clearly more consistent theoretically and more
realistic phenomenologically to keep the linear term of the potential,
as is done in Ref.~\cite{kni} but not in Ref.~\cite{ynd}.
On the other hand, both resonance and Green-function approaches of
Ref.~\cite{kni} effectively resum the contributions of soft multi-gluon
exchanges in the ladder approximation \cite{str,pri}.
This automatically includes the final-state interactions emphasized in
Ref.~\cite{ynd}.
However, we stress that all these methods are based on a non-relativistic
approximation.
There are additional contributions due to the exchange of hard gluons, which
give rise to sizeable reduction factors \cite{bar},
e.g., $(1-3C_F\alpha_s/\pi)$ in the case of $\Delta\rho$ \cite{kni}.

  Because of the more complicated potentials used in our two
approaches, we have to rely on numerical computations.  For  $m_t=\sqrt3m_Z$
and $\alpha_s(m_Z)=0.118  $, we find
\begin{eqnarray}
\label{eighta}
\delta(\Delta\rho)_{thr}&=&-(0.034\pm0.010)x_t
\qquad\mbox{(G.F.)},\\
\label{eightb}
\delta(\Delta\rho)_{thr}&=&-(0.042\pm0.013)x_t
\qquad\mbox{(res.)},
\end{eqnarray}
where we have included the 30\% error estimate given in Ref.~\cite{kni}.
The corresponding perturbative $\O(\alpha_s)$ contribution \cite{djo}
is, for $m_t=\sqrt3m_Z$,
\begin{eqnarray}
\delta(\Delta\rho)_{\alpha_s}&=&-{2\alpha_s(m_t)\over3\pi}
\left({\pi^2\over3}+1\right)x_t \nonumber\\
\label{nine}
&=&-0.0991\,x_t.
\end{eqnarray}
Unlike $\delta(\Delta\rho)_{thr}$, $\delta(\Delta\rho)_{\alpha_s}$
obtains important contributions arising from the
non-conserved vector and axial-vector currents associated with the
$W$-boson vacuum-polarization function.

It is apparent that the result for $\delta(\Delta\rho)_{thr}$ obtained in
the formulation of Ref.~\cite{ynd} (Eq.~(\ref{seven}))
is of the same magnitude as our two evaluations (Eqs.~(\ref{eighta}),
(\ref{eightb})).  It amounts to 49\% of the $\delta(\Delta\rho)_{\alpha_s}$
correction, while our results of Eqs.~(\ref{eighta}), (\ref{eightb}) correspond
to 34\% and 42\%, respectively.  Thus, it is somewhat larger than our
resonance calculation and significantly larger than our G.F.
result.  Part of the difference is due to the fact that we have
included a hard-gluon correction \cite{bar}, $(1-3C_F\alpha_s(M_n)/\pi)$
(cf.\ Eqs.~(4.2b, 4.3d) of Ref.~\cite{kni}), which has not been incorporated
into Eqs.~(\ref{three}), (\ref{four}).
Note that Eq.~(\ref{fivea}) must not be multiplied by this factor, since
only non-relativistic terms are subtracted in the integrand of that equation.
If this correction is applied to Eqs.~(\ref{three}), (\ref{four}),
$\delta(\Delta\rho)_{pole}$ becomes $-0.0245\,x_t$
and, rescaled to $\alpha_s(m_Z)=0.118$,
the overall result in the formulation of
Ref.~\cite{ynd} is $-0.0450\,x_t$, instead of Eq.~(\ref{seven}).
This is quite close to
our resonance calculation (Eq.~\ref{eightb}).  For the reasons explained in
Ref.~\cite{kni} (see also the discussion), for values of $m_t\ge130$~GeV
we have
expressed a preference for the G.F. approach.  On the other hand,
the three calculations amount to only 3.4\%--4.5\% of the
$\O(\alpha)$ contribution, $x_t$.

\begin{table} {TABLE~I. $t\bar t$ threshold effects on $\Delta\rho$
relative to the Veltman correction, $-100\times\delta(\Delta\rho)_{thr}/x_t$,
as a function of $m_t$.
The calculation based on Ref.~\cite{ynd} (total [6]) is compared with our
previous resonance (res.~[5]) and G.F. (G.F.~[5]) results.
For completeness, the contributions from below (pole~[6]) and above
(thr.~[6]) threshold are also displayed separately in the first case.
The hard-gluon correction is included in the three calculations
and the input value $\alpha_s(m_Z)=0.118$ is used.}\\[1ex]
\begin{tabular}{|c|c|c|c|c|c|} \hline
$m_t$ [GeV] & pole [6] & thr.~[6] & total [6] & res.~[5] & G.F. [5]\\ \hline
120.0 & 2.84 & 1.99 & 4.83 & 4.81 & 3.58 \\
140.0 & 2.73 & 1.92 & 4.65 & 4.43 & 3.43 \\
157.9 & 2.64 & 1.86 & 4.50 & 4.17 & 3.36 \\
180.0 & 2.55 & 1.79 & 4.35 & 3.91 & 3.35 \\
200.0 & 2.47 & 1.75 & 4.22 & 3.71 & 3.41 \\
220.0 & 2.41 & 1.70 & 4.11 & 3.53 & 3.51 \\
\hline
\end{tabular}
\end{table}

The above results have been obtained for $m_t=\sqrt3m_Z$, the value
employed in Ref.~\cite{ynd}.
In Table~I, we compare the calculation of $\Delta\rho$ using the formulation
of Ref.~\cite{ynd} with our own resonance and G.F. evaluations, over the
range 120~GeV${}\le m_t\le220$~GeV.
We have checked that the numbers given in the third and fourth columns of
Table~I do not change when we identify $v_0$ in Eq.~(\ref{fivea}) with the
zero of the integrand, i.e., the point where the resummation of
$(\pi C_F\alpha_s/v)^n$ terms matches the perturbative expression.
It is apparent from Table~I that the general features described above
hold over the large range 120~GeV${}\le m_t\le220$~GeV.
In fact, the three calculations are similar in magnitude.
Moreover, the formulation of Ref.~\cite{ynd} gives results somewhat
larger but quite close to our resonance calculation, the agreement
being particularly good at low $m_t$ values, where the resonance picture
is expected to work best.

\begin{table} {TABLE~II. Perturbative $\O(\alpha_s)$ and $t\bar t$ threshold
\cite{ynd} contributions to $\Delta\rho$ relative to the Veltman correction,
$-100\times\delta(\Delta\rho)_{\alpha_s}/x_t$ and
$-100\times\delta(\Delta\rho)_{thr}/x_t$, for
$m_t=\sqrt3m_Z$ as a function of $\mu_{pert}$.
The input value $\alpha_s(m_Z)=0.118$ is used.}\\[1ex]
\begin{tabular}{|c|c|c|c|} \hline
$\mu_{pert}$ & pert.~$\O(\alpha_s)$ & total [6] & sum \\ \hline
$m_t/2$ & 10.96 & 4.05 & 15.00 \\
$m_t$ & 9.91 & 4.50 & 14.42 \\
$2m_t$ & 8.87 & 5.18 & 14.05 \\
\hline
\end{tabular}
\end{table}

In order to illustrate the stability of the results with respect to a
change of the scale, $\mu_{pert}$, employed in the perturbative
$\O(\alpha_s)$ calculations, in Table~II we show the values of
$\delta(\Delta\rho)_{\alpha_s}$ (cf.\ Eq.~(\ref{nine})),
$\delta(\Delta\rho)_{thr}$, and
their sum for $m_t=\sqrt3m_Z$ and $\mu_{pert}=m_t/2,m_t,2m_t$.
Here $\delta(\Delta\rho)_{thr}$ is calculated on the basis of
Eqs.~(\ref{three}) and (\ref{fivea}), with $\alpha_s(m_t)$ replaced
by $\alpha_s(\mu_{pert})$ in the last term of Eq.~(\ref{fivea}), and
$v_0$ chosen as the zero of the integrand (the factor
$(1-v^2/3)/(1-v^2)^2$ discussed before Eq.~(\ref{seven}) is also appended).
We see that there are variations in $\delta(\Delta\rho)_{thr}$ of
$-10\%$ to 15\%, which are not particularly large.
Interestingly, they partly compensate the corresponding variations
in $\delta(\Delta\rho)_{\alpha_s}$.
Indeed, the overall QCD correction,
$\delta(\Delta\rho)_{\alpha_s}+\delta(\Delta\rho)_{thr}$,
which is the physically relevant quantity, changes by only $-3\%$ to 4\%,
a remarkably small variation.

\section{Bohr-atom estimate and other observations}

It is instructive to make a simple estimate of the threshold
effects on $ \Delta\rho$ by using an elementary Bohr-atom model of toponium
\cite{fad}. We have already employed this model to show that it leads to
values of $\left|R_{10}(0)\right|^2$ within 20\% of those obtained with the
Richardson potential \cite{fan}.
Now we want to apply it to illustrate the order of
magnitude of $\delta(\Delta\rho)_{thr}$.
Then, instead of Eq.~(\ref{three}), we have
\begin{equation}
\label{ten}
\delta(\Delta\rho)_{thr}=-x_t\,\pi C_F^3\sum_n{\alpha_s^3(k_n)\over n^3},
\end{equation}
where $k_n =C_F\alpha_s(k_n)m_t/(2n)$ is the momentum of the top quark in
the $nS$ orbital of
the Bohr-atom model.  We note that in this elementary estimate we
have evaluated $\alpha_s$ at scale $k_n$.  This is a simple generalization
of Ref.~\cite{fad}, where, for the ground state, $\alpha_s$ is evaluated at
$k_1=C_F\alpha_s(k_1)m_t/2$.
  In particular, for $m_t=\sqrt{3} m_Z$ and $\alpha_s(m_Z)=0.118$, we
 have $ k_1=16.7$~GeV and $\alpha_s(k_1)=0.159$.  In
the rough estimate of Eq.~(\ref{ten}), we have also disregarded the
continuum enhancement above threshold.  Because $k_n$ scales as $1/n$,
$\alpha_s(k_n)$ increases with $n$.  We have iteratively evaluated $k_n$
and $\alpha_s(k_n)$ up to $n=50$ and found
$\sum_{n=1}^{50}\alpha_s^3(k_n)/n^3=5.56\times10^{-3}$,
which is 1.38 times the $1S$ contribution. The sum converges rapidly;
the first 12 terms already yield a factor of 1.36.
Taking this to be an approximate estimate of the
enhancement factor due to the resonances with $n\ge 2$, and normalizing
Eq.~(\ref{ten}) relative to Eq.~(\ref{nine}), we have
\begin{equation}
\label{eleven}
{\delta(\Delta\rho)_{thr}\over\delta(\Delta\rho)_{\alpha_s}}
= 11.3 {\alpha_s^3 (k_1)\over\alpha_s(m_t)}.
\end{equation}
For $m_t=\sqrt3m_Z$ and $\alpha_s(m_Z)=0.118$, we have
$\alpha_s(m_t)=0.109$, $\alpha_s(k_1)=0.159$, and Eq.~(\ref{eleven})
gives 42\%,
which is rather close to the results obtained from the detailed
resonance approaches, i.e., Eqs.~(\ref{seven}) ,(\ref{eightb}) divided by
Eq.~(\ref{nine}). Although
Eq.~(\ref{eleven}) is a rough estimate, it allows us to understand why the
threshold effects on the $\rho$ parameter, although nominally of
$\O(\alpha_s^3)$, can be as large as $\approx40\%$ of the $\O(\alpha_s)$
contribution.  Two factors
are apparent: one is a large numerical coefficient, $\approx11$, and the
other is that the natural scale in the threshold contribution is
$k_1\ll m_t$, so that $\alpha_s(k_1)$ is considerably larger than
$\alpha_s(m_t)$.

A relevant question is how these effects compare with unknown
$\O(\alpha_s^2)$
contributions.  The leading $\O(\alpha_s)$ corrections to the $\rho$
parameter are $\approx10\%$ (cf.\ Eq.~(\ref{nine})).
If the same ratio holds between $\O(\alpha_s^2)$ and $\O(\alpha_s)$,
the threshold corrections we have discussed would be
roughly 3 to 4 times larger for $m_t\approx160$~GeV.
It is known that the use of
the running top-quark mass absorbs most of the $\O(\alpha_s)$ corrections
proportional to $m_t^2$ in the $\rho$ parameter and
$Z\to b\bar b$ amplitudes.
If this was a general feature of the perturbative expansion, one
could control the bulk of such contributions and the threshold
effects would neatly stand out.  However, it is impossible to
ascertain these features without detailed $\O(\alpha_s^2)$ calculations, which
are not available at present.  The effect of these threshold
corrections on the electroweak parameters are not particularly
large for $m_t\le220$~GeV.  For example, for $m_H=250$~GeV and
$m_t=(130,160,200)$~GeV we found \cite{fan}
shifts $\Delta m_W=-(42,55,77)$~MeV from the $\O(\alpha_s)$ contributions,
$\Delta m_W=-(14,19,27)$~MeV from threshold
effects evaluated with the resonance method, and
$\Delta m_W=-(10,16,25)$~MeV from threshold
effects evaluated in the G.F. approach.  The ratio of threshold to
$\O(\alpha_s)$ shifts in $m_W$ are similar but not identical to those we
encountered in $\Delta\rho$.  The reason is that the range of $m_t$
values of interest is not really in the asymptotic regime, and the
sub-leading $\O(\alpha_s)$ and threshold contributions to $\Delta r$
\cite{sir} (the relevant correction to calculate $m_W$) have somewhat
different $m_t$ dependences.  For example, for $m_t=160$~GeV and
$m_H=250$~GeV, the ratios for
$m_W$ shifts are 0.35 in the resonance approach and 0.29 in the G.F.
method \cite{fan}.
For $m_t=m_Z$ (220~GeV), the corresponding ratios amount to
0.25 (0.34) in the resonance approach and to 0.19 (0.34) in the
G.F. method, with the latter giving somewhat lower values below
$m_t=220$~GeV.

\section{Conclusions and discussion}

\begin{enumerate}
\item  We have pointed out that the conclusions of Ref.~\cite{ynd}
concerning the magnitude of the threshold effects discussed in Ref.~\cite{kni}
are without foundation.  They are based on the consideration of
a vector amplitude $\Pi(s)-\Pi(0)$, which, although very important in other
applications, has a very small effect in the analysis of Ref.~\cite{kni}.
This should be obvious because heavy particles of mass $m^2\gg s$
decouple in this amplitude.
\item  We have applied the analysis of $\im\Pi_{thr}(s^\prime)$ given in
Ref.~\cite{ynd} to study the threshold corrections to the
$\rho$ parameter for $m_t=\sqrt3m_Z$ and, after re-evaluating the quantities
involved, found a result that is similar in magnitude to those we
have reported previously.  Contrary to the conclusions of Ref.~\cite{ynd},
it is somewhat larger than our resonance calculation and
significantly larger than our Green-function (G.F.) evaluation.  In particular,
when hard-gluon contributions are included,
the result derived in the approach of Ref.~\cite{ynd} is
quite close to our own resonance calculation, well within the
theoretical errors quoted previously \cite{kni}.
As illustrated in Table~I, similar conclusions apply over the large range
120~GeV${}\le m_t\le220$~GeV.
We have also pointed out that the formulation of Ref.~\cite{ynd} leads,
for $m_t=\sqrt3m_Z$, to values of
$\delta(\Delta\rho)_{\alpha_s}+\delta(\Delta\rho)_{thr}$ which are remarkably
stable with respect to changes in the scale employed in the perturbative
calculation (see Table~II).
\item In our opinion, the fact that the formulation of Ref.~\cite{ynd},
when applied to $\Delta\rho$, gives results similar to ours (and, in fact,
quite close to our own resonance calculations), supports the notion that
these threshold effects can be reasonably estimated.
By the same token, it does not support recent claims of large ambiguities
in the threshold calculations \cite{hal}.
\item It is worthwhile to point out that, when the corrected values of
$\delta_{pole}$ and $\delta_{thr}$ obtained in the present paper are
applied, the threshold effects in $\Pi(s)-\Pi(0)$ amount to $\approx24\%$
of the perturbative $\O(\alpha_s)$ corrections evaluated at scale $m_t$.
Here $m_t=\sqrt3m_Z$ and $\alpha_s(m_Z)=0.115$ have been assumed
and, for simplicity, the hard-gluon correction is not included.
(The numerical evaluation reported in Ref.~\cite{ynd} gives, instead,
15\%).
These numbers are somewhat smaller than the effects we encountered in the
$\Delta\rho$ case: 34\% in the G.F. approach and 42\% in the resonance
framework (cf.\ Eqs.~(\ref{seven}), (\ref{eightb}) divided by
Eq.~(\ref{nine})).
However, we find it neither extraordinary nor unusual that
radiative-correction effects may vary by factors of 1.42, 1.75 or,
for that matter, 2.8, when applied to very different amplitudes.
In particular, the logic behind the conclusion of Ref.~\cite{ynd}
seems rather strange to us.
It is apparently based on the curious argument that a 15\% correction
in $\Pi(s)-\Pi(0)$ is considered to be very small and that, as a consequence,
a 34\% or 42\% effect in $\Delta\rho$ is intolerably large.
\item It is well known
that, for large $m_t$, the widths of the individual top quarks
become larger than the $1S$--$2S$ mass difference, so that the
bound-state resonances lose their separate identities and smear into a
broad threshold enhancement \cite{str}.  For this reason, we have expressed
a preference to use the resonance formulation for $m_t\le130$~GeV and the
G.F. approach for $m_t\ge130$~GeV \cite{kni}.
However, we do not regard either method as
``rigorous'' and, in fact, in Ref.~\cite{kni} we have assigned an estimated
30\% uncertainty to their evaluation.  In the analysis of
electroweak parameters, we have found that the resonance approach of
Ref.~\cite{kue} and the G.F. formulation of Ref.~\cite{str} lead to similar
results over a wide $m_t$ range, with the latter giving somewhat
smaller values for $m_t\le220$~GeV.
\item In order to make the relative size of
the threshold effects more readily understandable, we have
presented a simple estimate based on an elementary
Bohr-atom model of toponium (cf.\ Eq.~(\ref{eleven})),
and briefly commented on the possible
magnitude of $\O(\alpha_s^2)$ corrections.
\item We have not discussed here the non-perturbative contributions
connected with the existence of a gluon condensate, since they are
known to be exceedingly small in the case of the $t\bar t$ threshold
\cite{yak}; this has also been noticed in Ref.~\cite{ynd}.

\end{enumerate}

\bigskip
\centerline{\bf ACKNOWLEDGMENTS}
\smallskip
We would like to thank Paolo Gambino, Gustav Kramer,
Hans K\"uhn, Alfred Mueller, and Peter Zerwas for useful discussion.
This research was supported in part by the National Science
Foundation under Grant No.~PHY--9017585.


\begin{thebibliography}{99}
\bibitem{ren} In Ref.~\cite{ynd} there is a footnote claiming that an
expression in Ref.~\cite{kni} involving $|R_n(0)|^2$ is not
renormalization-group invariant.
This is not the case, as in our discussion $|R_n(0)|^2$ and the other terms
in that expression are evaluated at their physical scales (i.e.,
they do not depend on an arbitrary scale $\mu$).  The same would
occur in Eq.~(\ref{two}) if $\mu$ was identified with the relevant
physical scales, which are roughly the top-quark momenta in the
Bohr-atom orbitals.  In Eq.~(\ref{two}), the explicit $\mu$ dependence
compensates for the fact that $|\tilde R_{n0}^{(0)}|^2$ is
evaluated at scale $\mu$.
\bibitem{sca} Had we used an arbitrary scale $\mu$ in the calculation of the
last term in Eq.~(\ref{fivea}), this requirement would no longer be
satisfied.
\bibitem{wro} In Ref.~\cite{ynd}, the numerical value
$\delta_{thr}=5.4\times10^{-3}$ is reported, which is neither
consistent with Eq.~(\ref{fivea}) nor with Eq.~(\ref{six}).
\end{thebibliography}
\end{document}